\documentclass[%
aip,amsmath,amssymb,
reprint,
author-numerical,
hidelinks]{revtex4-1}
\usepackage[utf8]{inputenc}
\usepackage[T1]{fontenc}
\usepackage{amsmath}
\usepackage{amsfonts}
\usepackage{amssymb}
\usepackage[hidelinks]{hyperref}
\usepackage{empheq}
\usepackage{graphicx}
\usepackage{dcolumn}
\usepackage{mathptmx}

\usepackage{tikz}
\usetikzlibrary{shapes}
\usetikzlibrary{plotmarks}

\newcommand{\bv}{\mathbf{v}}
\newcommand{\bu}{\mathbf{u}}
\newcommand{\bE}{\mathbf{E}}
\newcommand{\bB}{\mathbf{B}}
\newcommand{\bF}{\mathbf{F}}

\newcommand{\bR}{\mathbf{R}}

\newcommand{\wn}{\widetilde{n}}

\begin{document}

\title{Generalized Impurity Pinch in Partially Magnetized Multi-Ion Plasma}
\date{\today}
\author{M. E. Mlodik} 
\email{mmlodik@pppl.gov}
\author{E. J. Kolmes}
\author{I. E. Ochs} 
\author{N. J. Fisch}
\affiliation{Department of Astrophysical Sciences, Princeton University, Princeton, New Jersey, USA, 08543 and \\
Princeton Plasma Physics Laboratory, Princeton, New Jersey, USA, 08540}

\begin{abstract}
	
	In a two-ion-species plasma with disparate ion masses, heavy ions tend to concentrate in the low-temperature region of collisionally magnetized plasma and in the high-temperature region of collisionally unmagnetized plasma, respectively. Moreover, collisional magnetization can be determined as the ratio of the light ion gyrofrequency to the collision frequency of light and heavy ion species, and the behavior of this effect in the intermediate regime of partially magnetized plasma is predominantly dependent on this Hall parameter. Multi-ion cross-field transport has been described before in the collisionally magnetized plasma regime, and generalized pinch relations, which describe densities of ion species in equilibrium in that plasma, are found in the literature. In this paper, the role of collisional magnetization and Larmor magnetization in multi-ion collisional transport is clarified and generalized pinch relations are extended to the partially magnetized regime, in which the ion Hall parameter may be small, as long as electrons remain collisionally magnetized. Equilibrium ion density profiles have the same dependence on external forces and on each other regardless of collisional magnetization of ions. The expansion of the range of validity of multi-ion collisional transport models makes them applicable to a wider range of laboratory plasma conditions. In particular, ion density profiles evolve sufficiently fast for radial impurity transport to be observable around stagnation on MagLIF, leading to expulsion of heavy ion impurities from the hotspot as long as plasma becomes sufficiently collisionally magnetized during the implosion.
	
\end{abstract}

\maketitle 

\section{Introduction.} 

In fusion devices, it is typically advantageous to concentrate fuel ions in the hot core of the plasma and to flush out impurities and fusion products  \cite{Hirshman1981, Redi1991, Braun2010, Ochs2018i, Ochs2018ii, Knapp2019, Gomez2019, Schmit2014}. This is true for a broad range of devices. 
For instance, a particular device where impurity transport is of interest\cite{Knapp2019, Gomez2019} is MagLIF\cite{Slutz2012,Gomez2014}, a magnetized Z-pinch device in Sandia. MagLIF features a cylinder of deuterium plasma, which is premagnetized by applying an axial $B$ field, heated in the center by a laser and compressed by applying a large current to the beryllium liner which envelops the fuel in order to reach fusion conditions. As such, there has been significant experimental effort to identify impurity mix properties\cite{Knapp2019,Harvey2018}, as well as technological developments such as preheat protocols to decrease the amount of impurity introduced to the fuel hotspot\cite{Knapp2019,Harvey2018,Gomez2019}. Nevertheless, multi-ion transport effects, i.e. impurity mix dynamics, have been overlooked in the currently available modeling of MagLIF implosions. Therefore, a relatively simple model of impurity dynamics is topical.

There is ample literature on similar plasmas, although they are different in at least one qualifier. Equlibrium in multi-ion magnetized plasmas in the absence of temperature gradients but in the presence of external forces was found by Kolmes et al\cite{Kolmes2018, Kolmes2020i}. Multi-ion plasma transport in the presence of temperature gradients, but in a very strong magnetic field (such that Hall parameters are much greater than 1) was studied by Ochs and Fisch\cite{Ochs2018i,Ochs2018ii}. More recently, the transport code MITNS has been developed\cite{Kolmes2021} to study evolution of multi-ion transport in collisionally magnetized plasmas. Transport in single-ion-species plasma with arbitrary Hall parameter of electrons was studied by Velikovich, Giuliani, and Zalesak\cite{Velikovich2015,Velikovich2019}. Multi-ion species transport in unmagnetized plasmas was studied by Kagan and Tang\cite{Kagan2012,Kagan2014}, as well as by Zhdanov and Alievskii\cite{Zhdanov,Alievskii1962,Zhdanov1980}. Some progress was also made in a study of multi-ion-species transport in a magnetized plasma in the presence of temperature gradients was made by Vekshtein et al\cite{Vekshtein1975}, although the authors did not have external forces and they had a specific configuration in mind, which is different from this paper.
The expansion of the range of applicability of muti-ion collisional transport models is also useful for plasma mass filters, which are designed to separate the components of a plasma according to mass \cite{Bonnevier1966, Lehnert1971, Hellsten1977, Krishnan1983, Geva1984, Bittencourt1987, Amoretti2006, Dolgolenko2017, Ochs2017iii, Yuferov2018, Gueroult2018} for a variety of applications \cite{Gueroult2014ii, Gueroult2015, Gueroult2018ii, Gueroult2019}. 

This paper provides a simplified model to describe multi-ion transport on devices like MagLIF. In particular, it includes conditions on ion densities in equilibrium in partially magnetized (large Hall parameter electrons, arbitrarily Hall parameter ions) multi-ion species plasma subjected to external forces or a temperature gradient, and clarifies what dimensionless parameters have the largest impact on multi-ion cross-field transport. Also, it suggests that collisional cross-field multi-ion transport is sufficiently fast to expel impurities from the hotspot during late stages of implosion. The effect is more profound if fuel plasma has collisionally magnetized ions, i.e. if the light ion gyrofrequency is much larger than the collision frequency between light and heavy ions.

The question of interest in this paper is how different species adjust to a change, such as change of temperature on the boundary (laser preheat after fuel magnetization) or the change of external potential (corresponding to Z-pinch compression and/or rotation). The problem is to find the force equilibrium state after that change happened, as well as to estimate how quickly ions adjust to that change relative to each other.

The paper is organized as follows. In Sec.~\ref{sec:Magnetization}, different regimes of multi-ion cross-field transport with respect to the strength of the magnetic field, and, in particular, the Hall parameter, are clarified. In Sec.~\ref{sec:Pinch}, generalized pinch relations, which relate ion densities in the force equilibrium, are derived in presence of both temperature gradients and external forces, such as the centrifugal force, in a two-ion-species plasma which has arbitrarily magnetized ions with large mass disparity $m_a \ll m_b$ as long as the Hall parameter of electrons is large, i.e. $\Omega_e/\nu_{ei} \gg 1$. Generalized pinch conditions turn out to depend predominantly on the Hall parameter $\Omega_a/\nu_{ab}$. 
In Sec.~\ref{sec:Timescales}, the multi-ion collisional cross-field transport timescale $\tau_{ab,eq}$ is derived in the low-$\beta$ limit.  The multi-ion cross-field transport timescale turns out to be sufficiently fast for impurity expulsion to occur at stagnation and at the later stages of implosion for MagLIF-relevant plasma parameters, even though it is much slower at the earlier stages of implosion.
In Sec.~\ref{sec:Summary} there is a summary of these results. In Sec.~\ref{sec:Discussion} there is a discussion of some other potential applications of these results. The procedure to find relevant transport coefficients is outlined in Appendix A.



\section{Plasma Magnetization.} 
\label{sec:Magnetization}

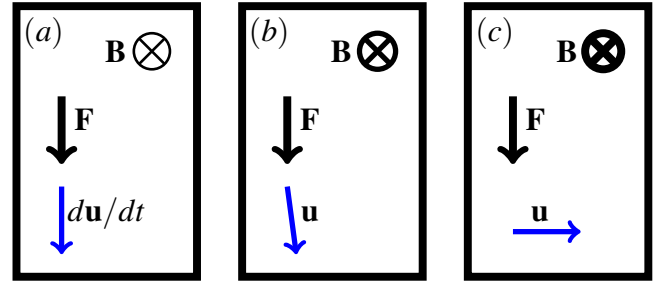
\begin{figure}
	\centering
	\begin{tikzpicture} [scale = 0.6]	
	\draw [line width = 3, fill = white] (-7, -3) rectangle (-3,3);
	
	\node [align = left] at (-6.4, 2.4) {\large $\textstyle{(a)}$};

	\draw [line width = 1, fill = white] (-4, 2) circle (0.4 cm); 
	\node [align = left] at (-4.8, 2) {\large $\textstyle{\mathbf{B}}$};
	
	\draw [line width = 1] ({-4+0.4*cos(135)}, {2+0.4*sin(135)}) to ({-4+0.4*cos(315)}, {2+0.4*sin(315)});
	\draw [line width = 1] ({-4+0.4*cos(45)}, {2+0.4*sin(45)}) to ({-4+0.4*cos(225)}, {2+0.4*sin(225)});

	\draw [line width = 3, ->, color = black] (-6, 1) to (-6, -0.5);
	\node [align = left] at (-6+0.5, 0.5) {\large $\bF$}; 	
	
	\draw [line width = 2, ->, color = blue] (-6, -1) to (-6, -2.5);
	\node [align = left] at (-6+1, -1.6) {\large $d\bu/dt$}; 	
	
	\draw [line width = 3, fill = white] (-2, -3) rectangle (2,3);
	
	\node [align = left] at (-1.4, 2.4) {\large $\textstyle{(b)}$};

	\draw [line width = 2, fill = white] (1, 2) circle (0.4 cm); 
	\node [align = left] at (0.2, 2) {\large $\textstyle{\mathbf{B}}$};
	
	\draw [line width = 2] ({1+0.4*cos(135)}, {2+0.4*sin(135)}) to ({1+0.4*cos(315)}, {2+0.4*sin(315)});
	\draw [line width = 2] ({1+0.4*cos(45)}, {2+0.4*sin(45)}) to ({1+0.4*cos(225)}, {2+0.4*sin(225)});

	\draw [line width = 3, ->, color = black] (-1, 1) to (-1, -0.5);
	\node [align = left] at (-1+0.5, 0.5) {\large $\bF$}; 	
	
	\draw [line width = 2, ->, color = blue] (-1, -1) to (-0.8, -2.5);
	\node [align = left] at (-0.9+0.4, -1.6) {\large $\bu$}; 
	
	
	\draw [line width = 3, fill = white] (3, -3) rectangle (7,3);
	
	\node [align = left] at (3.6, 2.4) {\large $\textstyle{(c)}$};

	\draw [line width = 3, fill = white] (6, 2) circle (0.4 cm); 
	\node [align = left] at (5.2, 2) {\large $\textstyle{\mathbf{B}}$};
	
	\draw [line width = 3] ({6+0.4*cos(135)}, {2+0.4*sin(135)}) to ({6+0.4*cos(315)}, {2+0.4*sin(315)});
	\draw [line width = 3] ({6+0.4*cos(45)}, {2+0.4*sin(45)}) to ({6+0.4*cos(225)}, {2+0.4*sin(225)});

	\draw [line width = 3, ->, color = black] (4, 1) to (4, -0.5);
	\node [align = left] at (4+0.5, 0.5) {\large $\bF$}; 	
	
	\draw [line width = 2, ->, color = blue] (4, -2) to (5.5, -2);
	\node [align = left] at (4.6, -1.6) {\large $\bu$};

	\end{tikzpicture}
	\caption{Classification of the response of species $s$ to an external force $\bF$ that is applied to a multi-ion-species plasma in a magnetic field. From left to right, in order of increase of the magnetic field: (a) Larmor and collisionally unmagnetized plasma $\rho_s/L \gg 1$; (b) Larmor magnetized, but collisionally unmagnetized plasma $\rho_s/L \ll 1, /\Omega_s/\nu_{ss'} \ll 1$; (c) Larmor and collisionally magnetized plasma $\rho_s/L \ll 1, \Omega_s/\nu_{ss'} \gg 1$. Partially magnetized plasma is an intermediate regime between (b) and (c).}
	\label{fig:magnetizationClassification}
\end{figure}

Consider a plasma slab in a homogeneous magnetic field $\mathbf{B} = B \hat{z}$ with species-dependent external potential $\Phi_s(y, t)$ with $\bF_s = - \nabla \Phi_s$ (and all other gradients) in the $\hat{y}$ direction. This plasma can be described by a multiple-fluid model \cite{Braginskii1965, Simakov2003}. 
The fluid momentum equation for species $s$ is
\begin{align}
m_s \frac{d\bu_s}{dt} = q_s \bE + q_s \bu_s \times \bB - \frac{\nabla p_s}{n_s} - \frac{\nabla \cdot \pi_s}{n_s} + \frac{\sum_{s'} \bR_{ss'}}{n_s} + \bF_s.
\label{eqn:momentumBasic}
\end{align}
Here $\bu_s$ is the flow velocity, $\pi_s$ is the traceless part of the pressure tensor of species $s$, and $\bR_{ss'} = \bR_{ss'}^u + \bR_{ss'}^T$ is the friction force, comprised of flow friction and  Nernst (``thermal") friction, between species $s$ and $s'$ \cite{Hinton, HelanderSigmar}. 
In the limit where $m_s \ll m_{s'}$ and the Hall parameter of the light species is large, $\bR^T_{ss'} = 3 n_s \nu_{ss'} \hat{b} \times \nabla T_s / 2 \Omega_s$. 
The magnetic field enters the momentum equation in two ways: explicitly in the Lorentz force and implicitly in transport coefficients. The magnetization can be understood in two ways: in the $\rho_s /L$ sense, as the smallness of the gyroradius compared to the characteristic length scale of perpendicular dynamics, and in the $\Omega_s/\nu_s$ sense, as a ratio of gyrofrequency to collision frequency. Plasmas which exhibit these two types of magnetization can be called \textit{Larmor magnetized} ($\rho_s/L \ll 1$) and \textit{collisionally magnetized} ($\Omega_s/\nu_s \gg 1$), respectively. Also, plasmas such that $\Omega_s/\nu_s \sim 1$ can be called \textit{partially magnetized}. In principle, there can be a few different choices of a particular collision frequency or a gyrofrequency, but later in this paper it is found that in the case of light ion species $a$ and heavy ion species $b$ the relevant Hall parameter is $\Omega_a/\nu_{ab}$.
In order to see how magnetization impacts Eq.~(\ref{eqn:momentumBasic}), compare the inertia term and the pressure term:
\begin{align}
\frac{|m_s (\bu_s \cdot \nabla) \bu_s|}{|q_s \bu_s \times \bB|} \sim \frac{u_s^2 / L}{q_s u_s B / m_s} \sim \frac{u_s}{\Omega_s L} \sim \frac{u_s}{u_{th,s}} \frac{\rho_s}{L}.
\label{eqn:momentumTermsRatio}
\end{align}
The ratio above shows the velocity response of plasma species to an external force: in $\rho_s/L \ll 1$ case the velocity is proportional to the external force, while in $\rho_s/L \gg 1$ the acceleration is proportional to the external force. Classification of these regimes is shown in Fig.~\ref{fig:magnetizationClassification}.
In a sense, the right-hand side of Eq.~(\ref{eqn:momentumTermsRatio}) can be understood as an analogue to the Reynolds number in the Navier-Stokes equation: it is a dimensionless parameter which delineates qualitatively different kinds of system behavior depending on which term in the momentum equation is dominant. 
In Larmor magnetized plasma, the Lorentz force is much larger than the ion inertia. 
In collisionally magnetized plasma (Fig.~\ref{fig:magnetizationClassification}c), the Lorentz force is much larger than the ion-ion friction force. Therefore, plasma responds to the external force by having $\bF \times \bB$ drift in the direction perpendicular to the external force $\bF$, and the drift velocity is such that the Lorentz force and the force $\bF$ are balanced. In collisionally unmagnetized plasma (Fig.~\ref{fig:magnetizationClassification}b) the ion-ion friction force is much larger than the Lorentz force. Therefore, the external force $\bF$ is balanced by the ion-ion friction force. Since the direction of the ion-ion friction force and the difference in the ion flow velocities are close to each other, the difference in the ion flow velocities is pointing in the similar direction to the external force $\bF$.
The Lorentz force and the ion-ion friction force are comparable in partially magnetized plasma. In many systems of interest, such as Z-pinches, plasma is Larmor magnetized but collisionally unmagnetized. Similarly, timescales are such that $1/\Omega_s \cdot \partial/\partial t \ll 1$ holds true. From now on, only Larmor magnetized plasmas are considered in the paper.

Rearranging Eq.~(\ref{eqn:momentumBasic}), the following expression is obtained:
\begin{align}
\bu_s \times \hat{\mathbf{b}} =  - & \frac{\bE}{B} + \frac{1}{\Omega_s} \frac{d \bu_s}{dt} + \frac{\nabla p_s}{m_s n_s \Omega_s} + \frac{\nabla \cdot \pi_s}{m_s n_s \Omega_s} \nonumber\\ - &  \frac{\sum_{s'} \bR_{ss'}}{m_s n_s \Omega_s} - \frac{\bF_s}{q_s B}.
\label{eqn:momentumCrossed}
\end{align}
In order to have the closure of Eq.~(\ref{eqn:momentumCrossed}), expressions for $\pi_s$ and $\bR_{ss'}$ are needed. Note, however, that the viscosity $\pi_s$ is small compared to other terms so it affects only long-term behavior. As such, in many cases it can be ignored. As far as the friction force $\bR_{ss'}$ goes, it can be found from the expression
\begin{align}
\bR_{ss'} = \int d^3 \bu ~m_s \bu C_{ss'} \left(f_s, f_{s'}\right).
\end{align}
Here $C_{ss'} (f_s,f_{s'})$ is a collision operator which describes collisions between species $s$ and $s'$. Note that the friction force depends on the distribution functions $f_s$. In the case of $\epsilon = \rho_{L,i}/L$ being small, a perturbative expansion of distribution functions in powers of $\epsilon$ can be performed around non-perturbed Maxwellian with zero mean velocity, $f_s = f_{s0} + f_{s1} + ...$, as long as the plasma is sufficiently collisional to enforce $f_{s1} \ll  f_{s0}$.
Then $f_{s1}$ satisfies the following equation \cite{Hinton}:
\begin{align}
	& \sum_{s'} \big[  C_{ss'} (f_{s1}, f_{s'0}) + C_{ss'} (f_{s0}, f_{s'1}) \big] + \Omega_s \frac{\partial f_{s1}}{\partial \xi}  = \nonumber\\ &  \bu \cdot \left[ \left(\frac{\nabla p_s}{p_s} - \frac{q_s \bE}{T_s} - \frac{\bF_s}{T_s}\right) + \left( \frac{u^2}{u_{th,s}^2} - \frac{5}{2}\right) \frac{\nabla T_s}{T_s} \right] f_{s0}.
\label{eqn:f1}
\end{align}
Here $\xi$ is the gyrophase. The friction force, inertia term, and viscosity do not enter the right-hand side of Eq.~(\ref{eqn:f1}) as they are ordered down as $\mathcal{O}(\epsilon), \mathcal{O}(\epsilon^2),\mathcal{O}(\epsilon^2)$, respectively. 

Note that $f_{s1}$ depends on collisions with all other species and on same-species collisions too, as long as collisions are sufficiently frequent compared to Larmor gyration. Therefore, if there are multiple unmagnetized or partially magnetized ion species, the friction force $\bR_{ss'}$ depends not only on the behavior of species $s$ and $s'$, but on the behavior of all species. In the magnetized case, Eq.~(\ref{eqn:f1}) is solved by Hinton\cite{Hinton}. The unmagnetized case is solved by Zhdanov and Alievskii\cite{Zhdanov,Alievskii1962}, and it has been successfully applied by Kagan and Tang\cite{Kagan2012,Kagan2014}. In the partially magnetized case, however, it is a challenging task to find even a closed form expression for friction force. Nevertheless, simplifications can be made if: (1) there are two ion species with disparate masses or (2) all ion species except for one are in trace quantities. The aim of this paper is to elucidate case 1.

In the case of two ion species with disparate masses $m_a \ll m_b$, the expression for the friction force can be derived in the following way, as long as electrons are collisionally magnetized, i.e. $\Omega_e/\nu_{ei} \gg 1$. The distortion of the light ion distribution function $f_a$ is much larger than the distortion of $f_b$. As such, magnetization is determined entirely by the light ion species. The most general form of the components of the friction force between light and heavy species which are perpendicular to magnetic field is\cite{Braginskii1965,Epperlein1986}
\begin{align}
\bR_{ba} &= - m_b n_b \nu_{ba} \big[ \alpha_{\perp,ba} \left( \bu_b- \bu_a \right) + \alpha_{\wedge,ba} \left( \bu_b - \bu_a \right) \times \hat{\mathbf{b}} \big] \nonumber\\&+ \beta_{\perp,ba} n_a \nabla_{\perp} T_a - \beta_{\wedge,ba} n_a \hat{\mathbf{b}} \times \nabla T_a.
\label{eqn:frictionForce}
\end{align}
Curiously, in the case of large mass disparity between ions, transport coefficients for ion-ion friction (Eq.~(\ref{eqn:frictionForce})) in two-ion-species plasma are the same as transport coefficients for electron-ion friction in single-ion-species plasma up to the following substitution: $\Omega_e / \nu_{ei} \to \Omega_a / \nu_{ab} $, $\widetilde{Z} \to n_b Z_b^2 / (n_a Z_a^2)$ (see derivation in Appendix \ref{section:FrictionForceExpression}). Coefficients for the electron-ion friction force in the case of arbitrary magnetization have been found by Epperlein and Haines\cite{Epperlein1986}.
In the limit of collisionally magnetized plasma (suppressing indices $ba$ attached to the transport coefficients), 
$\Omega_a/\nu_{ab} \to \infty $: $\beta_{\perp} / \beta_{\wedge} \to 0$, $\beta_{\wedge} \to 3/2 \cdot \nu_{ab} / \Omega_a$, $\alpha_{\perp} \to 1$.
In the limit of collisionally unmagnetized plasma,
$\Omega_a/\nu_{ab} \to 0$: $\beta_{\wedge} / \beta_{\perp} \to 0$, $\beta_{\perp} \to const$, $\alpha_{\perp} \to const$.
$\alpha_{\wedge} / \alpha_{\perp} $ goes to $0$ in both limits, and never exceeds $0.2$. 

In partially magnetized plasma, the ion-ion friction force has unusual terms, such as the component perpendicular both to the magnetic field and the flow velocity difference (term $\alpha_{\perp,ba} \left( \bu_b - \bu_a \right) \times \hat{\mathbf{b}} $ in Eq.~(\ref{eqn:frictionForce})). Therefore, it is instructive to provide an intuitive explanation of why this component exists in the first place. In the single-particle picture, its origin can be attributed to the following observation. In plasma with ion Hall parameter $\sim 1$, ions' trajectories are arcs of a Larmor orbit, interrupted by collisions. The length of these arcs is inversely proportional to the collision frequency. Therefore, if there are more collisions on the one side of the orbit than another, there is going to be net motion in the perpendicular direction. In comparison, in collisionally magnetized plasma a particle makes many gyrations between successive collisions, so the effect of collisions on the arc length is averaged out. In collisionally unmagnetized plasma, particles' trajectories are close to straight lines, so there is almost no preference in the motion perpendicular to the flow velocity difference. Therefore, this component of the friction force is significant only in partially magnetized plasma regime.

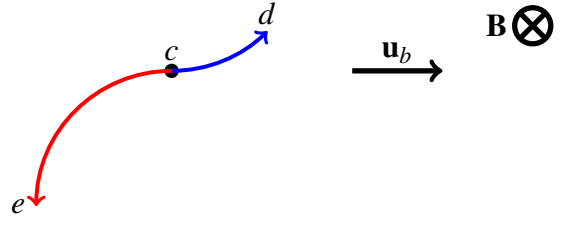
\begin{figure}
	\centering
	\begin{tikzpicture} [scale = 0.6]
	
	\draw [line width = 2, fill = white] (5, 2) circle (0.4 cm); 
	\node [align = left] at (4.2, 2) {\large $\textstyle{\mathbf{B}}$};
	
	\draw [line width = 2] ({5+0.4*cos(135)}, {2+0.4*sin(135)}) to ({5+0.4*cos(315)}, {2+0.4*sin(315)});
	\draw [line width = 2] ({5+0.4*cos(45)}, {2+0.4*sin(45)}) to ({5+0.4*cos(225)}, {2+0.4*sin(225)});

	\draw [line width = 2, fill = black] (-3, 1) circle (0.1); 
	\node [align = left] at (-3, 1+0.4) {\large $c$}; 	
	
	\draw [line width = 1.5, ->, color = blue] (-3,1) arc (270:315:3);
	\node [align = left] at ({-3+3*cos(315)}, {1+3+3*sin(315)+0.4}) {\large $d$}; 	

	\draw [line width = 1.5, ->, color = red] (-3,1) arc (90:180:3);	
	\node [align = left] at ({-3+3*cos(180)-0.4}, {1-3+3*sin(180)}) {\large $e$}; 	

	\draw [line width = 2, ->, color = black] (1, 1) to (3, 1);
	\node [align = left] at (2, 1.4) {\large $\bu_b$}; 	
	
	\end{tikzpicture}
	\caption{Single-particle picture of the origin of non-collinearity of the ion-ion friction force and their flow velocity difference in partially magnetized plasma. Consider two particles of species $a$ moving with the opposite velocities and starting at the same point $c$. The particle which has the velocity in the same direction as flow velocity of species $b$ (blue trajectory) has smaller relative velocity difference than the particle which has the velocity in the opposite direction (red trajectory). As collision frequency in plasma decreases dramatically with the increase of relative velocity difference, the particle on the blue trajectory is going to collide much faster (at point $d$) than the particle on the red trajectory (at point $e$). Therefore, the arc $ce$ is much larger than the arc $cd$.}
\end{figure}

Another way to get an intuition about the role of collisional magnetization is to look at the direction of the distortion of distribution function $f_{s1}$ using Eq.~(\ref{eqn:f1}) in a plasma where there is an imbalance of $\hat{x}$-direction components of flow velocities of two species. To see that, consider a case of light species $a$ and heavy species $b$. Then the distribution function of the heavy species $f_b$ is relatively unaffected by the species $a$. Moreover, the spread of $f_b$ is much smaller than the spread of the distribution function of light species $f_a$. Therefore, light species essentially see the delta-function distribution of heavy species. Collisions with heavy species essentially provide a force that pushes light species distribution toward the mean velocity of heavy species. Moreover, it also means that the behavior of $f_{b1}$ can be ignored and all the essential physics are concentrated in the distortion of the distribution function of the light species $f_{a1}$.

In collisionally magnetized plasma, the reaction of $f_{a1}$ is going to be in $\hat{y}$ direction, perpendicular both to $\hat{x}$, the direction of the flow velocity imbalance and to $\hat{z}$, the direction of the magnetic field. Then this $\hat{y}$-directed distortion provides $\hat{x}$-directed Lorentz force, balancing the unlike-species friction force.

In collisionally unmagnetized plasma, like-species collisions push the distribution function of the light species back to Maxwellian, and the Lorentz force is comparatively too weak. Therefore, $f_{a1}$ is in the $\hat{x}$-direction.

In partially magnetized plasma, both mechanisms of pushing the distribution function of the light species back to Maxwellian are equally important. Therefore, $f_{a1}$ is stretched out in both $\hat{x}$ and $\hat{y}$ directions. The stretching of $f_{a1}$ in both directions creates an asymmetry in average relative velocity difference between particles moving in the positive and the negative $\hat{y}$-direction, which in turn results in the asymmetry of the collisional drag. Summed over all the light-species particles, this is seen as the component of the friction force perpendicular to the direction of flow imbalance. Note that both the presence of magnetic field (creating a component of  distortion in $\hat{y}$ direction) and the collision frequency dependence on relative velocity is essential for this component of the friction force to appear. Also, in the collisionally magnetized plasma case the distortion of $f_a$ in $\hat{y}$-direction is much smaller than in the partially magnetized plasma case, so the $\alpha_{\perp,ba} \left(\bu_b - \bu_a \right) \times \hat{b}$ component of the friction force is much smaller, too.

\section{Equilibrium and Generalized Pinch Relations.}
\label{sec:Pinch}

Consider a magnetized multiple-ion species slab of plasma which is subjected to a slow ($\partial / \partial t \ll \nu_{ss'}, \Omega_s$) change in the applied potential or temperature in the direction perpendicular to the magnetic field. Then cross-field dynamics are governed by collisional transport. 

On the fast transport timescales (compared to the viscous transport timescale) momentum equation for species $s$ (Eq.~(\ref{eqn:momentumBasic})) has the following form in the equilibrium:
\begin{gather}
0 = q_s \bE + q_s \bu_s \times \bB - \frac{\nabla p_s}{n_s} + \frac{\sum_{s'} \bR_{ss'}}{n_s} + \bF_s.
\label{eqn:momentumEquilibrium}
\end{gather}
Suppose all the gradients are in the $x$-direction and the magnetic field is in the $z$-direction. Then in equilibrium, $u_{sx} = 0$ for all species $s$ and Eq.~(\ref{eqn:momentumEquilibrium}) becomes
\begin{gather}
0 = q_s E_y + \frac{\sum_{s'} R_{ss',y}}{n_s}.
\label{eqn:momentumEquilibriumY}
\end{gather}
Summing Eq.~(\ref{eqn:momentumEquilibriumY}) over all species, the condition that $E_y = 0$ is obtained (no inductive $\bE$ field in the equilibrium).
Therefore, there are $s$ equilibrium conditions: $\sum_s R_{ss',y} = 0$. These equilibrium conditions can be shown to provide relations between densities of plasma constituents, which can be called \textit{generalized pinch relations}. At much longer timescales, viscosity plays a role, and plasma evolves to the state of thermodynamic equilibrium. 

\subsection{Transitivity of Generalized Pinch Relations.}

In general, the force equilibrium conditions are $\sum_{s'} R_{ss',y} =0$. However, in two important cases: (a) no more than three species or (b) uniform temperature $\nabla T = 0$, they can be replaced with pairwise friction force cancellation $R_{ss',y} =0$. If there are three or fewer species, the $\sum_{s'} R_{ss',y} = 0$ condition together with $R_{ss',y} = - R_{s's,y}$ provides a sufficient number of constraints to uniquely determine $R_{ss',y}$, and $R_{ss',y} = 0$ is the equilibrium. If there is no temperature gradient, $R_{ss',y} = R_{ss',y}^u \propto u_{s'y} - u_{sy}$, so in equilibrium all flow velocities $\bu_s$ are the same, and therefore $R_{ss',y} = 0$.

In a system with $N$ species in force equilibrium, the zero-friction-force conditions will provide $N-1$ independent constraints. The last constraint, which is needed to determine density profiles of all species, depends on timescales and $\beta$; see Sec.~\ref{sec:Timescales} for more details.

In plasma of arbitrary magnetization without temperature gradients, the maximum-entropy approach\cite{Kolmes2020i} gives the following result:
\begin{gather}
\bigg( n_a e^{\Phi_a / T} \bigg)^{1/Z_a} \propto \bigg( n_{b} e^{\Phi_{b}/T} \bigg)^{1/Z_{b}}. \label{eqn:phiPinch}
\end{gather}
In collisionally magnetized plasma this condition is equivalent to setting the friction force between each pair of species to zero, see Ref. \cite{Spitzer1952, Taylor1961ii, Braginskii1965, Kolmes2018}. Also, in a collisionally magnetized plasma there are results about pinch relations with temperature gradients\cite{Rutherford1974, Hinton1974, Hinton1976, Wade2000,  HelanderSigmar, Dux2004, Helander2017, Newton2017}. However, this is the first time that Eq.~(\ref{eqn:phiPinch}) has been derived for plasma that is not in the collisionally magnetized regime. In fact, the analysis in this section provides the first new, nontrivial test of the maximum-entropy derivation of Eq.~(\ref{eqn:phiPinch}) since that approach was presented in Ref.\cite{Kolmes2020i}. A partially magnetized plasma fulfills all of the requirements given in Ref.\cite{Kolmes2020i} for a system to satisfy Eq.~(\ref{eqn:phiPinch}), and we have shown here (independently of any entropy considerations) that indeed Eq.~(\ref{eqn:phiPinch}) is satisfied by the equilibrium for this system.

Transitivity is a convenient property of generalized pinch relations, as it makes the equilibrium state of any pair of species independent of the properties of the other species which comprise the plasma.

\subsection{Generalized Pinch Relations in Two-Ion-Species Plasma.}

In the presence of temperature gradients, generalized pinch relations become less tractable as the expression for the friction force becomes complicated and the maximum-entropy principle is no longer directly applicable.
However, in the two-ion-species plasma case, equilibrium requires pairwise friction force cancellation: $R_{ab,y} = R_{ae,y} = R_{be,y} = 0$.

Different species also can move across magnetic field lines at different rates. Therefore, in some cases\cite{Mlodik2020,Kolmes2021}, there is  separation of timescales in collisional transport. In particular, in collisionally magnetized plasma ions equilibrate between each other faster than with electrons. Such fast transport timescale equilibrium (in which generalized pinch relations are satisfied for a subset of species, but not for all of them) will have $R_{ab,y} = 0$, but not necessarily $R_{ae,y} = R_{be,y} = 0$. The question of transport timescales is addressed in more detail in Sec.~\ref{sec:Timescales} and \ref{sec:Summary}.

In the case of two ion species with disparate masses $m_a \ll m_b$ and collisionally magnetized electrons (i.e. $\Omega_e/\nu_{ei} \gg 1$), generalized pinch conditions can be found explicitly using the expression for the friction force (Eq.~(\ref{eqn:frictionForce})).
In equilibrium, the conditions $R_{ab,y}=0$ and $u_{ax} = u_{bx} = 0$ can be combined with Eq.~(\ref{eqn:frictionForce}) as
\begin{gather}
-m_b n_b \nu_{ba} \alpha_{\perp,ba} \left( u_{by} - u_{ay} \right) - \beta_{\wedge,ba} n_a T_a' = 0.
\label{eqn:momentumExplicitY}
\end{gather}
Here and later in the paper, $f' = \partial f/\partial x$ for any physical quantity $f$. Define
\begin{gather}
\beta_{\star,ba} = \beta_{\perp,ba} + \frac{\beta_{\wedge,ba} \alpha_{\wedge,ba}}{\alpha_{\perp,ba}}.
\end{gather}
Then Eq.~(\ref{eqn:momentumExplicitY}) can be used to find that in equilibrium
\begin{gather}
R_{ba,x} = \beta_{\star,ba}  n_a T_a'.
\end{gather}
Note that in collisionally magnetized plasma the friction force decays quickly ($\propto (\Omega_{a}/\nu_{ab})^{-5/3}$, see Ref.\cite{Epperlein1986} and Appendix \ref{section:FrictionForceExpression}). Since electrons are substantially more magnetized than ions, electron-ion friction force is much smaller than ion-ion friction force in equilibrium as long as electrons are collisionally magnetized.

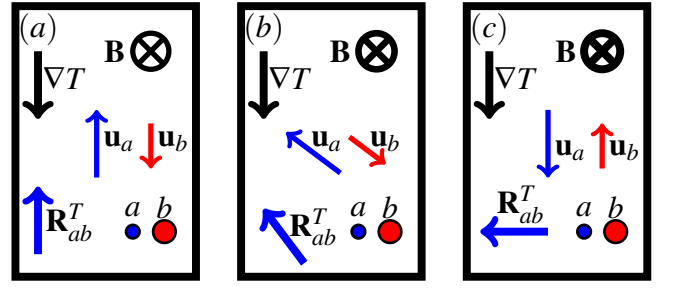
\begin{figure}
	\centering
	\begin{tikzpicture} [scale = 0.6]
	
	
	\draw [line width = 3, fill = white] (-7, -3) rectangle (-3,3);
	
	\node [align = left] at (-6.5, 2.5) {\large $\textstyle{(a)}$};

	\draw [line width = 2, fill = white] (-4, 2) circle (0.4 cm); 
	\node [align = left] at (-4.8, 2) {\large $\textstyle{\mathbf{B}}$};
	
	\draw [line width = 2] ({-4+0.4*cos(135)}, {2+0.4*sin(135)}) to ({-4+0.4*cos(315)}, {2+0.4*sin(315)});
	\draw [line width = 2] ({-4+0.4*cos(45)}, {2+0.4*sin(45)}) to ({-4+0.4*cos(225)}, {2+0.4*sin(225)});

	\draw [line width = 3, ->, color = black] (-6.5, 2) to (-6.5, 0.5);
	\node [align = left] at (-6.5+0.6, 1.4) {\large $\nabla T$}; 	
	
	\draw [line width = 3, ->, color = blue] (-6.5, -2.5) to (-6.5, -1);
	\node [align = left] at (-6.5+0.7, -1.8) {\large $\bR^T_{ab}$}; 	
	
	\draw [line width = 2, ->, color = blue] (-5.2, -0.8) to (-5.2, 0.7);
	\node [align = left] at (-5.2+0.5, 0) {\large $\bu_a$}; 
	
	\draw [line width = 2, ->, color = red] (-4, 0.4) to (-4, -0.6);
	\node [align = left] at (-4+0.5, 0) {\large $\bu_b$}; 
	
	\draw [line width = 1, fill = blue] (-4.4, -2) circle (0.15); 
	\node [align = left] at (-4.4, -1.5) {\large $\textstyle{a}$};
	
	\draw [line width = 1, fill = red] (-3.7, -2) circle (0.25); 
	\node [align = left] at (-3.7, -1.45) {\large $\textstyle{b}$};
	
	\draw [line width = 3, fill = white] (-2, -3) rectangle (2,3);
	
	\node [align = left] at (-1.5, 2.5) {\large $\textstyle{(b)}$};

	\draw [line width = 2.5, fill = white] (1, 2) circle (0.4 cm); 
	\node [align = left] at (0.2, 2) {\large $\textstyle{\mathbf{B}}$};
	
	\draw [line width = 2.5] ({1+0.4*cos(135)}, {2+0.4*sin(135)}) to ({1+0.4*cos(315)}, {2+0.4*sin(315)});
	\draw [line width = 2.5] ({1+0.4*cos(45)}, {2+0.4*sin(45)}) to ({1+0.4*cos(225)}, {2+0.4*sin(225)});

	\draw [line width = 3, ->, color = black] (-1.5, 2) to (-1.5, 0.5);
	\node [align = left] at (-1.5+0.6, 1.4) {\large $\nabla T$}; 	
	
	\draw [line width = 3, ->, color = blue] (-0.6, -2.7) to (-1.5, -1.5);
	\node [align = left] at (-0.4, -1.9) {\large $\bR^T_{ab}$}; 	
	
	\draw [line width = 2, ->, color = blue] (0.2, -0.7) to (-1, 0.2);
	\node [align = left] at (-0.1, 0) {\large $\bu_a$}; 
	
	\draw [line width = 2, ->, color = red] (0.4, 0.1) to (1.2, -0.5);
	\node [align = left] at (1.2, 0) {\large $\bu_b$}; 
	
	\draw [line width = 1, fill = blue] (0.6, -2) circle (0.15); 
	\node [align = left] at (0.6, -1.5) {\large $\textstyle{a}$};
	
	\draw [line width = 1, fill = red] (1.3, -2) circle (0.25); 
	\node [align = left] at (1.3, -1.45) {\large $\textstyle{b}$};

	
	\draw [line width = 3, fill = white] (3, -3) rectangle (7,3);
	
	\node [align = left] at (3.5, 2.5) {\large $\textstyle{(c)}$};

	\draw [line width = 3, fill = white] (6, 2) circle (0.4 cm); 
	\node [align = left] at (5.2, 2) {\large $\textstyle{\mathbf{B}}$};
	
	\draw [line width = 3] ({6+0.4*cos(135)}, {2+0.4*sin(135)}) to ({6+0.4*cos(315)}, {2+0.4*sin(315)});
	\draw [line width = 3] ({6+0.4*cos(45)}, {2+0.4*sin(45)}) to ({6+0.4*cos(225)}, {2+0.4*sin(225)});

	\draw [line width = 3, ->, color = black] (3.5, 2) to (3.5, 0.5);
	\node [align = left] at (3.5+0.6, 1.4) {\large $\nabla T$}; 	
	
	\draw [line width = 3, ->, color = blue] (4.8, -2) to (3.3, -2);
	\node [align = left] at (4.2, -2+0.6) {\large $\bR^T_{ab}$}; 	
	
	\draw [line width = 2, ->, color = blue] (4.8, 0.7) to (4.8, -0.8);
	\node [align = left] at (4.8+0.5, -0.2) {\large $\bu_a$}; 
	
	\draw [line width = 2, ->, color = red] (6, -0.6) to (6, 0.4);
	\node [align = left] at (6+0.5, -0.2) {\large $\bu_b$}; 
	
	\draw [line width = 1, fill = blue] (5.6, -2) circle (0.15); 
	\node [align = left] at (5.6, -1.5) {\large $\textstyle{a}$};
	
	\draw [line width = 1, fill = red] (6.3, -2) circle (0.25); 
	\node [align = left] at (6.3, -1.45) {\large $\textstyle{b}$};

	\end{tikzpicture}
	\caption{Thermal force and impurity pinch. In collisionally unmagnetized plasma (a), where ion Hall parameter is small $\Omega_a/\nu_{ab} \ll 1$, both thermal force and flow velocity are collinear with the temperature gradient. As a result, heavy ions tend to concentrate in a hotter region of plasma. In collisionally magnetized plasma (c), where ion Hall parameter is large $\Omega_a/\nu_{ab} \gg 1$, thermal force is perpendicular to the direction of temperature gradient. However, ion flow velocity is perpendicular to the thermal force in this case. As a result, heavy ions tend to concentrate in a colder region of plasma. In partially magnetized plasma (b) , where $\Omega_a/\nu_{ab} \sim 1$, both thermal force and flow velocity are not collinear to the temperature gradient or to each other. In all cases, collisional cross-field transport due to temperature gradient conserves charge locally, i.e. $Z_a \bu_a + Z_b \bu_b = 0$.}
	\label{fig:thermalForce}
\end{figure}

The momentum equation in $x$-direction is
\begin{gather}
0 = q_s E_x + q_s u_{sy} B - \frac{p_s'}{n_s} + \frac{\sum_{s'} R_{ss',x}}{n_s} + F_{sx}.
\end{gather}
Adopting $\Omega_s = q_s B / m_s$, $F_s^{\star} = F_s - p_s'/n_s$,
\begin{gather}
u_{sy} = - \frac{E_x}{B} - \frac{F_s^{\star}}{m_s \Omega_s} - \frac{\sum_{s'} R_{ss',x}}{m_s n_s \Omega_s}.
\label{eqn:velocityY}
\end{gather}
Combined with Eq.~(\ref{eqn:momentumExplicitY}):
\begin{align}
& - \frac{F_b^{\star}}{m_b \Omega_b} - \frac{R_{ba,x} + R_{be,x}}{m_b n_b \Omega_b} \nonumber\\ &  + \frac{F_a^{\star}}{m_a \Omega_a} + \frac{R_{ab,x} + R_{ae,x}}{m_a n_a \Omega_a} + \frac{\beta_{\wedge,ba} n_a T_a' }{m_b n_b \nu_{ba} \alpha_{\perp,ba}}= 0.
\end{align}
Together with $m_a n_a \nu_{ab} = m_b n_b \nu_{ba}$ and the fact that electron-ion friction force can be ignored as long as $\Omega_e/\nu_{ei} \gg 1$, 
\begin{align}
\frac{F_a^{\star}}{m_a \Omega_a} + \frac{\beta_{\wedge,ba} \Omega_a}{\alpha_{\perp,ba} \nu_{ab}} \frac{T_a' }{m_a \Omega_a} - \frac{R_{ba,x}}{m_a n_a \Omega_a} \left(1 + \frac{Z_a n_a}{Z_b n_b} \right)= \frac{F_b^{\star}}{m_b \Omega_b}.
\label{eqn:pinchNoElectronsUnsimplified}
\end{align}
Eq.~(\ref{eqn:pinchNoElectronsUnsimplified}) can be rewritten as
\begin{align}
\frac{1}{Z_b} \left(\frac{p_b'}{n_b} - F_{bx} \right) = \frac{1}{Z_a} \left(\frac{p_a'}{n_a} - F_{ax} + \lambda T_a' \right),
\label{eqn:pinchGeneralized}
\end{align}
where 
\begin{gather}
\lambda = \beta_{\star,ba} \left(1 + \frac{Z_a n_a}{Z_b n_b}\right) - \frac{\beta_{\wedge,ba} \Omega_a}{\alpha_{\perp,ba} \nu_{ab}}
\label{eqn:temperatureScreening}
\end{gather}
is the temperature screening coefficient. $\lambda$ describes the effect of the thermal force on the equilibrium density profiles. In collisionally unmagnetized plasma, $\lambda$ is positive, while in collisionally magnetized plasma $\lambda$ is negative. The way in which thermal force affects multi-ion transport in partially magnetized plasma is shown in Fig. \ref{fig:thermalForce}.

Eq.~(\ref{eqn:pinchGeneralized}) can be called the \textit{generalized pinch relation in partially magnetized plasma}. It provides a constraint on densities of ion species $a$ and $b$ in force equilibrium, as long as $m_a \ll m_b$ and $\Omega_e/\nu_{ei} \gg 1$. The generalized pinch relation makes it possible to analyze characteristics of impurity transport in plasma with arbitrary collisional magnetization of ions. 

In limiting cases, Eq.~(\ref{eqn:pinchGeneralized}) reduces to the forms known in the literature.
In collisionally magnetized plasma, $\lambda = -3/2$ and Eq.~(\ref{eqn:pinchGeneralized}) is reduced to (see Ref.\cite{Spitzer1952, Taylor1961ii, Braginskii1965, Kolmes2018,Ochs2018i})
\begin{gather}
\frac{1}{Z_b} \left(\frac{p_b'}{n_b} - F_{bx} \right) = \frac{1}{Z_a} \left(\frac{p_a'}{n_a} - F_{ax} - \frac{3}{2}T_a' \right).
\label{eqn:pinchMagnetized}
\end{gather}

In isothermal plasma it is reduced to Eq.~(\ref{eqn:phiPinch}), as could be expected from the maximum-entropy principle\cite{Kolmes2020i}.

The collisionally unmagnetized plasma limit of Eq.~(\ref{eqn:pinchGeneralized}) can also be found ($\beta_{\perp,ba} > 0$):
\begin{gather}
\frac{1}{Z_b} \left(\frac{p_b'}{n_b} - F_{bx} \right) = \frac{1}{Z_a} \left(\frac{p_a'}{n_a} - F_{ax} + \beta_{\perp,ba} \left(1 + \frac{Z_a n_a}{Z_b n_b}\right) T_a' \right).
\label{eqn:pinchUnmagnetized}
\end{gather}

In order to find out the dependence of the heavy species' density profile on temperature and the meaning of the temperature screening coefficient $\lambda$, consider the case of no external forces acting on the plasma: $F_a = F_b = 0$. Then Eq.~(\ref{eqn:pinchGeneralized}) becomes
\begin{align}
\frac{1}{Z_b} \frac{p_b'}{n_b} = \frac{1}{Z_a} \left(\frac{p_a'}{n_a} + \lambda T_a' \right),
\end{align}
It can be reorganized to see the dependence on density and temperature gradients separately (using equation of state $p_s = n_s T_s$):
\begin{gather}
\frac{n_b'}{n_b} = \frac{Z_b}{Z_a} \frac{T_a}{T_b} \frac{n_a'}{n_a} + \left( \lambda + 1  \right) \frac{Z_b T_a}{Z_a T_b} \frac{T_a'}{T_a} - \frac{T_b'}{T_b}.
\label{eqn:pinchNoForce}
\end{gather}
If species $a$ and $b$ have the same temperature $T_a = T_b = T$, Eq.~(\ref{eqn:pinchNoForce}) can be simplified further.
\begin{gather}
\frac{n_b'}{n_b} = \frac{Z_b}{Z_a} \frac{n_a'}{n_a} + \left[ \left( \lambda + 1  \right) \frac{Z_b }{Z_a } - 1 \right] \frac{T'}{T}.
\label{eqn:pinchNoForceSimplified}
\end{gather}
It is clear from Eq.~(\ref{eqn:pinchNoForceSimplified}) that the density gradient of light ion species $a$ and the temperature gradient act provide two independent contributions to the density gradient of heavy species $b$. 
As follows from Eq.~(\ref{eqn:pinchNoForceSimplified}), heavy species are expelled from high-$T$ regions when plasma is collisionally magnetized and $\lambda \to -3/2$. 
However, they are concentrated in high-$T$ regions instead when plasma is collisionally unmagnetized. Eq.~(\ref{eqn:temperatureScreening}) determines the boundary between these qualitatively different types of plasma behavior. Note, however, that in general the sign of the effect changes at the point $\lambda = Z_a/Z_b - 1$ and not at the point $\lambda = 0$, as the temperature gradient is also present in $p_s'$ terms, while $\lambda$ describes only the contribution of the thermal force. It turns out that the temperature screening coefficient $\lambda$ depends predominantly on the ion Hall parameter $\Omega_a/\nu_{ab}$, since transport coefficients in  Eq.~(\ref{eqn:temperatureScreening}) vary significantly with the change of $\Omega_a/\nu_{ab}$ and comparatively weakly with the $Z_{eff} = n_b Z_b^2 / (n_a Z_a^2)$ (see Ref.\cite{Epperlein1986} and Appendix~\ref{section:FrictionForceExpression} for details). Therefore, $\Omega_a/\nu_{ab}$ is the criterion to determine collisional magnetization of plasma.

Note that the dependence of heavy species' density on light species' density and on external forces, such as the centrifugal force, is the same regardless of whether plasma is collisionally magnetized or not. Aside from that, this section provides an extensive overview of the impact of collisional magnetization on temperature gradient-driven multi-ion transport.

\section{Timescales in Partially Magnetized Plasma.}
\label{sec:Timescales}

Sec.~\ref{sec:Pinch} describes the equilibrium state of multi-ion-species plasma. However, it is of significant interest whether the knowledge of relationship between ion density profiles can also be extended to laboratory plasmas which are dynamically evolving. In this section, there is an estimate of the timescale of relative relaxation of ion density profiles toward the equilibrium, as well as an example of laboratory plasma where this timescale is sufficiently fast to make the multi-ion transport effects observable.

\subsection{Timescale of Ion-Ion Force Equilibration.}

In general, as plasma evolves due to cross-field transport, all densities and temperatures are subject to change. Therefore, it is not easy to describe the resulting evolution of plasma. However, in the special case of low-$\beta$ isothermal plasma ($T_s = T = const$ across the plasma for all species $s$; $\beta \ll 1$), which is slightly out of equilibrium and is subjected to small forces ($F_s L \ll T$ for all species $s$, where $L$ is a length scale of plasma), it is possible to track the plasma behavior as a function of time. In this particular case, Eq.~(\ref{eqn:pinchGeneralized}) prescribes that densities are close to being uniform, i.e. $\left|\nabla n_s\right| / n_s \ll 1 / L$. Then, in order to linearize the momentum and continuity equations, assume $n_s = n_{s0} + \wn_s$, $\bu_s = \bu_{s0} + \widetilde{\bu}_s$, where $n_{s0}$ and $\bu_{s0}$ are the equilibrium density and flow velocity of species $s$, respectively. Moreover, $\nabla n_s /n_s \approx \nabla \wn_s/ n_{s0}$. The friction force can also be linearized in the same way: $\bR_{ss'} = \bR_{ss'0} + \widetilde{\bR}_{ss'}$. Then in partially magnetized plasma Eq.~(\ref{eqn:momentumCrossed}) becomes
\begin{align}
\widetilde{\bu}_s \times \hat{\mathbf{b}} =  \frac{T \nabla \wn_s}{m_s n_{s0} \Omega_s} -   \frac{\sum_{s'} \widetilde{\bR}_{ss'}}{m_s n_{s0} \Omega_s}.
\label{eqn:momentumLinearized}
\end{align}
Continuity equation:
\begin{gather}
\frac{\partial \wn_s}{\partial t} + \nabla \cdot \left( n_{s0} \widetilde{\bu}_s \right) = 0.
\label{eqn:continuity}
\end{gather}
The expression for the perturbation of the friction force between light species $a$ and heavy species $b$ can be inferred from Eq.~(\ref{eqn:frictionForce}):
\begin{align}
\widetilde{\bR}_{ba} = - m_b n_b \nu_{ba} \big[ \alpha_{\perp,ba} \left( \widetilde{\bu}_b- \widetilde{\bu}_a \right) + \alpha_{\wedge,ba} \left( \widetilde{\bu}_b- \widetilde{\bu}_a \right) \times \hat{\mathbf{b}} \big].
\label{eqn:frictionLinearized}
\end{align}
Define transport timescale $\tau_{ab,eq}$ as a characteristic timescale of plasma solely due to friction between species $a$ and $b$, ignoring interactions with other species. Then Eqs.~(\ref{eqn:momentumLinearized}) and (\ref{eqn:continuity}) can be combined to get the fact that frictional transport is ambipolar:
\begin{gather}
Z_a \wn_a + Z_b \wn_b = 0.
\end{gather}
Moreover, Eqs.~(\ref{eqn:momentumLinearized}) and (\ref{eqn:frictionLinearized}) can be combined (omitting index $ab$ in $\alpha_{\perp}$ and $\alpha_{\wedge}$) as
\begin{align}
& \frac{T}{m_a n_{a0} \Omega_a} \nabla \wn_a - \frac{T}{m_b n_{b0} \Omega_b} \nabla \wn_b = - \left( \widetilde{\bu}_b - \widetilde{\bu}_a \right) \times \hat{\mathbf{b}} \nonumber \\  + & \left( \frac{\nu_{ab}}{\Omega_a}  + \frac{\nu_{ba}}{\Omega_b} \right) \left( \alpha_{\perp} \left(\widetilde{\bu}_b - \widetilde{\bu}_a\right) + \alpha_{\wedge} \left( \widetilde{\bu}_b - \widetilde{\bu}_a\right)  \times \hat{\mathbf{b}} \right).
\end{align}
If all the gradients are in one direction, these equations can be combined to get a diffusion equation
\begin{gather}
\frac{\partial \wn_a}{\partial t} + \nabla \cdot \left( D_{ab} \nabla \wn_a \right) = 0,
\label{eqn:diffusion}
\end{gather}
where the diffusion coefficient is
\begin{gather}
D_{ab} = \frac{\alpha_{\perp} \left( \nu_{ab} \rho_a^2 + \nu_{ba} \rho_b^2 \right)}{\left[1 - \alpha_{\wedge} \left( \frac{\nu_{ab}}{\Omega_a} + \frac{\nu_{ba}}{\Omega_b} \right) \right]^2 + \alpha_{\perp}^2 \left( \frac{\nu_{ab}}{\Omega_a} + \frac{\nu_{ba}}{\Omega_b} \right)^2 }.
\label{eqn:diffusionCoefficient}
\end{gather}
Here $\rho_s^2 = T/(m_s \Omega_s^2)$ is the gyroradius of species $s$.
In the limit of collisionally magnetized plasma ($\Omega_a/\nu_{ab} \gg 1, \Omega_b/\nu_{ba} \gg 1$), Eq.~(\ref{eqn:diffusionCoefficient}) reduces to 
\begin{gather}
D_{ab} =  \nu_{ab} \rho_a^2 + \nu_{ba} \rho_b^2.
\label{eqn:diffusionCollMagnetized}
\end{gather} 
In the limit of collisionally unmagnetized plasma, $\alpha_{\wedge} \ll \alpha_{\perp}$, so the diffusion coefficient is similar to what could be expected ($D \sim \nu  \lambda_{mfp}^2  $):
\begin{gather}
D_{ab} = \frac{\nu_{ab} \rho_a^2 + \nu_{ba} \rho_b^2}{  \alpha_{\perp} \left( \frac{\nu_{ab}}{\Omega_a} + \frac{\nu_{ba}}{\Omega_b} \right)^2 }.
\end{gather}
In the limit of trace heavy impurities, $n_b \to 0$, $\Omega_a/\nu_{ab} \to \infty$, so the diffusion coefficient attains limit value
\begin{gather}
D_{ab} = \frac{\nu_{ba}}{\nu_{ba}^2 + \Omega_b^2} \frac{T}{m_b}.
\end{gather}

Eq.~(\ref{eqn:diffusion}) can be solved by spectral decomposition. In a cylinder the timescale of equilibration of the lowest mode due to radial ion-ion transport is
\begin{gather}
\tau_{ab,eq} = \frac{1}{j_{1,1}^2} \frac{r^2}{D_{ab}}.
\label{eqn:timescalesCylinder}
\end{gather}

Here $1/j_{1,1}^2 = 0.06811...$ is a geometric factor (the square of the first zero of the first-order Bessel function); this geometric factor should be $1/\pi^2$ in slab geometry. In the limit of collisionally magnetized plasma ($\Omega_a/\nu_{ab} \gg 1, \Omega_b/\nu_{ba} \gg 1$), it is the same as in Ref.\cite{Mlodik2020} up to a geometric factor. 

Eq.~(\ref{eqn:timescalesCylinder}) has been found in the limit of low-$\beta$ isothermal plasma. In high-$\beta$ plasma, by analogy to neutral gas\cite{Geyko2016,Kolmes2020ii}, there can be other ways in which plasma evolves to the equilibrium, such as magnetosonic waves. Another way to introduce complications to the dynamics would be to include evolution of $B$ and $T$. Nevertheless, Eq.~(\ref{eqn:timescalesCylinder}) provides an idea about how long it takes for densities of ion species to adjust relative to each other once some change has been applied to the system.
Another feature to note in Eqs.~(\ref{eqn:diffusionCoefficient}) and (\ref{eqn:timescalesCylinder}) is that $\tau_{ab,eq}$ approaches a constant value in the limit of trace impurity species $b$.

Another way to interpret transport timescale $\tau_{ss',eq}$ is that it is the timescale of relaxation of the friction force $R_{ss'}$ toward zero, after which generalized pinch condition Eq.~(\ref{eqn:pinchGeneralized}) for species $s,s'$ can be used. If one of the species $s,s'$ is electrons, temperature screening coefficient is $\lambda = - 3/2$ in the range of applicability of the results in this paper.

\subsection{Timescales of Z-pinch Implosion Around Stagnation.}

One important area of applicability of generalized pinch relations is impurity transport in magnetized Z-pinch experiments, such as MagLIF, where plasma is Larmor magnetized, but not necessarily collisionally magnetized. MagLIF implosions start with magnetization and preheat of fuel, forming a hotter and less dense region of plasma in the center, called the hotspot. Then the implosion itself happens, compressing the fuel until stagnation. 

\begin{figure}
	\centering
	\begin{tikzpicture} [scale = 0.8]
	
	\draw [line width = 2, ->, color = black] (-4, -4) to (4, -4);
	\node [align = left] at (3.5, -3.5) {\large $\log(r_0/r)$}; 
	
	\draw [line width = 2, ->, color = black] (-4, -4) to (-4, 3);
	\node [align = left] at (-2.6, 2.8) {\large $\log(\tau_{ab,eq})$}; 
	
	\draw[line width = 1, - , color = black] (-3.5,2.5) to (-0.5,-2.5);
	
	\draw[line width = 1, - , color = black] (-0.5,-2.5) to (2.5,-1.5);
	
	\draw[line width = 1, dashed, ->, color = black] (-0.5, -4.5) to (-0.5,-2.5);
	\node [align = left] at (-0.2, -4.7) { $\nu_{ab}/\Omega_a + \nu_{ba}/\Omega_b \sim 1$}; 
	
	\draw[line width = 1, dashed, color = black] (-4, -0.5) to (4,-0.5);
	\node [align = left] at (3, -0.2) { $\tau_{ab,eq} = \tau_{imp}$}; 
	
	\node[align = left] at (-1.4, 1.1) { $\sim (r_0/r)^{-10/3}$}; 
	
	\node[align = left] at (2.8, -1.8) { $\sim (r_0/r)^{2/3}$}; 
	
	\end{tikzpicture}
	\caption{Evolution of radial multi-ion transport timescale in a typical magnetized Z-pinch implosion. Initially, plasma is collisionally unmagnetized and radial particle transport is too slow compared to the duration of the implosion. Then collisional transport becomes much faster, until sum of ion Hall parameters becomes $\sim 1$. Afterward, transport timescale starts to increase. Increase of Hall parameter during the implosion also means that impurities tend to be expelled from the hotspot more and more as Z-pinch approaches stagnation phase.}
	\label{fig:implosionTimescales}
\end{figure}
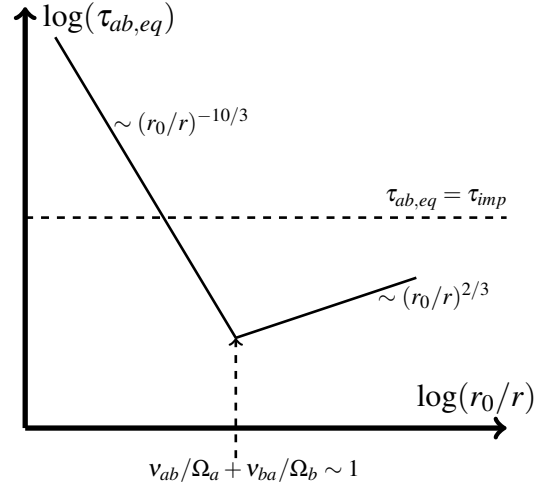

Suppose for simplicity that the implosion is a radial metric compression with convergence ratio $C = r_0/r$ of a cylindrical plasma column which has adiabatic index $\gamma$. Then plasma parameters in the fuel have the following scaling: $n \propto C^2$, $T \propto n^{\gamma - 1} \propto C^{2\left(\gamma - 1\right)}$, $B \propto C^2$. Gyrofrequencies and collisional frequencies have $\Omega_s \propto C^2$ and $\nu_{ss'} \propto n/T^{3/2} \propto C^{5-3\gamma}$ scaling, respectively. Dimensionless parameters that characterize magnetization in various ways have the following scalings: $\beta \propto nT/B^2 \propto C^{2\left(\gamma - 2\right)}$, $\Omega_a/\nu_{ab} \propto C^{3\left(\gamma-1\right)}$, $\rho_a/r \propto T^{1/2}B^{-1}r^{-1} \propto C^{\gamma-2}$. As such, the plasma becomes more collisionally magnetized during the implosion. In the collisionally unmagnetized plasma limit $\tau_{ab,eq} \propto r^2 \nu_{ab} / v_{th}^2 \propto C^{-5\left(\gamma-1\right)}$; in the collisionally magnetized plasma limit $\tau_{ab,eq} \propto r^2 / \left(\nu_{ab} \rho_b^2 \right) \propto C^{\gamma-1}$.

Collisional radial transport of heavy impurities in a Z-pinch implosion can be understood as the following (see also Fig.~\ref{fig:implosionTimescales}). Initially, even after the axial magnetic field is applied to the fuel, plasma is still collisionally unmagnetized ($\Omega_a/\nu_{ab} \ll 1$) and transport timescale $\tau_{ab,eq}$ is too slow compared to the implosion time for collisional multi-ion transport to have much impact. Then, as implosion progresses, $\tau_{ab,eq}$ drops dramatically ($\propto C^{-10/3}$ if $\gamma=5/3$). Therefore, impurities start to be drawn into the hotspot due to temperature gradients and expelled from the hotspot due to density gradients.
Then, when plasma becomes partially magnetized and, afterwards, collisionally magnetized, the generalized pinch effect flips sign. Therefore, heavy ion impurities get expelled from the hotspot both due to temperature and density gradients, as hotspot is both hotter and less dense than the surrounding plasma. Moreover, ion-ion transport timescale is sufficiently fast for the impurities to actually get expelled. After that, $\tau_{ab,eq}$ starts to increase. 

This analysis could be applied to  a general class of magnetized Z-pinch configurations.
For example, assume plasma parameters similar to the recent high-performance shot z3040 on Z machine, as described in Ref.\cite{Gomez2020}. In that shot, initially $\rho R = 0.68~mg/cm^3$, $r=2.3~mm$, $B = 10~T$. If the impurity concentration is $n_{Be} = 0.1 n_{D}$ and the ion temperature is $T_i = 100~eV$, then $\nu_{D,Be}/\Omega_D = 71.6$ and $\tau_{BeD,eq} \approx 9~\mu s$, i.e. radial multi-ion collisional transport is too slow. It would be even slower if impurity concentration is smaller or fuel is colder. Therefore, collisional radial transport cannot be expected to be prominent in the beginning of the implosion. Also, $\Omega_e/\nu_{ei} \approx 0.5$ in this plasma, so the plasma is initially out of the regime of applicability of the model in this paper. However, during the compression, the plasma becomes partially magnetized and then collisionally magnetized, so the impurity transport can be described by the model presented in this paper in the later stages of compression.

In particular, if plasma parameters at stagnation $BR = 0.2~MG\cdot cm$, $r = 60~\mu m$ (which corresponds to $B = 3.3~kT$), fuel density is $\rho R = 0.4~g/cm^3$, $T_i  = 2.6~keV$ are considered, both ion Hall parameter and transport timescale are much more favorable. In particular, if there is $n_{Be} = 0.05 n_{D}$, then $\Omega_D/\nu_{D,Be} = 0.78$ and $\tau_{BeD,eq} = 2.1~ns$. If $n_{Be} = 0.01 n_{D}$, then $\Omega_D/\nu_{D,Be} = 3.75$ and $\tau_{BeD,eq} = 3.7~ns$. The multi-ion radial transport timescale is comparable to the burn time ($\sim 2~ns$) and is much smaller than the implosion timescale ($\sim 100~ns$). As such, significant impurity transport can be expected to occur. Hall parameter for electrons is $\Omega_e/\nu_{ei} \approx 28$ in this case, well in the range of applicability of the transport model used in this paper.

This paper describes what happens to impurities when $\Omega_e/\nu_{ei} \gg 1$ and $m_a \ll m_b$. These conditions are satisfied in later stages of implosion and around stagnation in the current experiments. However, this is precisely when impurity transport matters the most. Moreover, the preliminary results presented here are promising: both density and temperature gradients tend to expel impurities from the hotspot as long as $\Omega_a/\nu_{ab} \gg 1$, and transport timescale $\tau_{ab,eq}$ is sufficiently short for this to happen.

\section{Summary.} 
\label{sec:Summary}

Multi-ion cross-field transport has been studied in the literature in the limit of collisionally magnetized ions. 
There is also literature on the role of collisional magnetization of electrons in the single-ion species plasma, as well as on multi-ion transport in unmagnetized plasma.
However, the intermediate regime of partially magnetized plasma has largely not been addressed until now.

This paper provides a rigorous classification of different magnetization regimes and their role in multi-ion transport. 
Moreover, it gives the conditions on density profiles in equilibrium in the case of partially magnetized plasma, 
thereby extending multi-ion collisional transport models to a new regime of applicability.
Temperature gradient dependence of generalized pinch relations allows to recover both the collisionally magnetized limit, where heavy ions are expelled from the high-$T$ region of plasma, and the collisionally unmagnetized limit, where heavy ions are drawn in the high-$T$ region of plasma. Moreover, in the case of light ion species $a$ and heavy ion species $b$ the sign of this effect depends almost exclusively on the Hall parameter $\Omega_a/\nu_{ab}$, while like-species Hall parameter $\Omega_a/\nu_{aa}$  changes only the magnitude of the effect. To some extent, the results in this paper can be generalized to the case of multiple heavy impurity species, as long as they are in trace quantities, such that their collision frequencies with one another are negligible compared to their collision frequencies with bulk ions.

The extension of the parameter space where multi-ion collisional transport is understood expands the range of applicability of transport models to more laboratory plasmas.
In particular, the results in this paper are promising in terms of understanding impurity transport in Z-pinches, such as the MagLIF experiment.
More generally, multi-ion collisional transport leads to a significant radial expulsion of heavy ion impurities if two conditions are satisfied: (i) Hall parameter $\Omega_a/\nu_{ab} \gtrsim 1$ being large at stagnation; (ii) transport timescale $\tau_{ab,eq}$ being comparable to or faster than the length of the stagnation and late implosion phases. 

\section{Discussion.}
\label{sec:Discussion}

In low-$\beta$ collisionally magnetized plasma, there is separation of timescales between ion-ion transport timescale $\tau_{ab,eq}$ and electron-ion transport timescale $\tau_{ei,eq}$: $\tau_{ab,eq} \ll \tau_{ei,eq}$. Therefore, plasma exhibits curious effects on $\tau_{ab,eq}$, such as charge incompressibility and the heat pump effect\cite{Mlodik2020,Kolmes2021}. More generally, in partially magnetized plasma, when electrons are still collisionally magnetized, but ions are only partially magnetized, the ratio of cross-field transport timescales is the following:
\begin{gather}
\frac{\tau_{ei,eq}}{\tau_{ab,eq}} = \frac{D_{ab}}{D_{ei}}.
\label{eqn:timescalesSeparation}
\end{gather}
Eqs.~(\ref{eqn:diffusionCoefficient}) and (\ref{eqn:timescalesSeparation}) can be combined to identify the parameter space where there is  separation of transport timescales. In particular, in the case of trace heavy ion species $b$ and bulk light ion species $a$, timescales separation exists if
\begin{gather}
\left(\frac{\nu_{ab}}{\Omega_a} + \frac{\nu_{ba}}{\Omega_b} \right)^2 \lesssim \frac{\nu_{ab}}{\nu_{ae}}.
\end{gather}
Timescales separation between multi-ion transport and electron-ion transport potentially makes it possible to observe charge incompressibility and the heat pump effect even in partially magnetized plasma.

Even though magnetized Z-pinches are providing an excellent testbed for the application of the theory of multi-ion collisional transport in partially magnetized plasma, they do not exhaust the potential upside of the knowledge of heavy ion density profiles. In particular, the tendency of heavy ions to concentrate in the region of $\Omega_a/\nu_{ab} \sim 1$ due to a temperature gradient can be used for ion separation using compression or heating. Another potential application of the multi-ion transport models in partially magnetized plasma is to identify the range of applicability of the idea of introducing trace ions to increase the fuel concentration in the hotspot, as shown by Kim et al in Ref.\cite{Kim2021}. In Ref.\cite{Kim2021}, H ions were used on Omega experiment (unmagnetized) to increase the concentration of DT fuel in the hotspot and reduce demixing, as heavier ions tend to concentrate in high-$T$ region of plasma in that case. However, the prescription would be the opposite in the partially magnetized or collisionally magnetized plasma. Therefore, in these plasmas the use of H ions would lead to the opposite result of fuel concentration in low-$T$ region of plasma, and heavy ions should be used instead. The model presented in this paper makes it possible to identify the threshold when this effect changes sign.

A potential extension of this work is to include finite-$\beta$ effects on the time-dependent evolution of multi-ion transport, which would allow more accurate quantitative predictions of the evolution of the density profiles, complementing the knowledge about the multi-ion equilibrium in this paper. Another interesting extension would be to include the role of particle fluxes on the generalized pinch relations\cite{Mitra2020} to the partially magnetized plasma regime.

\acknowledgments{The authors thank R. Gueroult, S. Davidovits, T. Rubin, and E. Mitra for useful conversations. This work was supported by Cornell NNSA 83228-10966 [Prime No. DOE (NNSA) DE-NA0003764] and by NSF-PHY-1805316.}

\providecommand{\noopsort}[1]{}\providecommand{\singleletter}[1]{#1}%

\appendix
\section{Expression for Ion-Ion Friction Force through Electron-Ion Transport Coefficients.}
\label{section:FrictionForceExpression}
One of the ways to simplify the calculation of density profiles is to use the same expression for the ion-ion friction force  as for the electron-ion friction in a single-ion-species plasma. Consider plasma that has two ion species $a$ and $b$, such that $m_a \ll m_b$. 
\begin{gather}
\bR_{ab} = \int d^3\bv~ m_a \bu C_{ab} (f_a, f_b).
\label{eqn:frictionForceCopy}
\end{gather}
The friction force $\bR_{ab}$ comes from distortion of distribution functions $f_a$ and $f_b$. But in the $m_a \ll m_b$ case, species $a$ does not see a distortion of $f_b$ because the thermal spread of $f_b$ is much smaller than the spread of $f_a$. Therefore, only distortion of $f_a$ contributes to the friction force. Suppose that the distortion of $f_a$ from Maxwellian distribution is small, $f_{a1} \ll f_{a0}$.
Then $f_a$  satisfies the following relation (see Eq.~(173) in Hinton\cite{Hinton}, or Eq.~(\ref{eqn:f1})):
\begin{align}
C_{aa} (f_a, f_a) + C_{ab} (f_a, f_b) + C_{ae} (f_a, f_e) + \Omega_a \frac{\partial f_{a1}}{\partial \xi} = \nonumber \\ \bu \cdot \left[ \left(\frac{\nabla p_a}{p_a} - \frac{q_a \bE}{T_a} \right) + \left( \frac{u^2}{u_{th,a}^2} - \frac{5}{2}\right) \frac{\nabla T_a}{T_a} \right] f_{a0}.
\end{align}
Here $u_{th,a} = \sqrt{2 T_a/m_a}$. Since $C_{ae} \propto \nu_{ae}$, it is a small correction relative to the other terms on the LHS. As long as $C_{ae}$ term can be ignored,
\begin{align}
 & C_{aa} (f_a, f_a) + C_{ab} (f_a, f_b) + \Omega_a \frac{\partial f_{a1}}{\partial \xi} = \nonumber \\  & \bu \cdot \left[ \left(\frac{\nabla p_a}{p_a} - \frac{q_a \bE}{T_a} \right) + \left( \frac{u^2}{u_{th,a}^2} - \frac{5}{2}\right) \frac{\nabla T_a}{T_a} \right] f_{a0}.
\label{eqn:fa1}
\end{align}
Here $C_{aa}$ is a same-species collision operator, and $C_{ab}$ is a collision operator between light and heavy species (e.g. pitch-angle scattering operator). In comparison, in a single-ion-species plasma equation for $f_e$ is
\begin{align}
& C_{ee} (f_e, f_e) + C_{ei} (f_e, f_i) + \Omega_e \frac{\partial f_{e1}}{\partial \xi} = \nonumber \\ &  \bu \cdot \left[ \left(\frac{\nabla p_e}{p_e} - \frac{q_e \bE}{T_e} \right) + \left( \frac{u^2}{u_{th,e}^2} - \frac{5}{2}\right) \frac{\nabla T_e}{T_e} \right] f_{e0}.
\label{eqn:fe1}
\end{align}
The RHS as a function of $u/u_{th,s}$ is the same up to index change $e \to a$. Therefore, if the coefficients are such that all the terms 
on the LHS of Eqs.~\ref{eqn:fa1} and \ref{eqn:fe1} are proportional, distortion of distribution function satisfies the same equation (since operators $C_{aa}$ and $C_{ee}$ are similar, the same is true for $C_{ab}$ and $C_{ei}$). Then the solution for $f_{a1}$ is also the same as the solution for $f_{e1}$, or at least these solutions are proportional to each other. Therefore, the corresponding friction forces $R_{ab}$ and $R_{ei}$ are also proportional to each other.

Then $C_{ab}$ can be split into terms ${}^{1}C_{ab}$ and ${}^{2}C_{ab}$ that depend on $f_{a1}$ and $\bu_b$, respectively (using Eq.~(121) in Ref.\cite{Hinton} as the expression for $C_{ab}$), and collision operators and $f_a$ can be rewritten in dimensionless form (adorned by tilde), moving ${}^{2}C_{ab}$ to the other side:
\begin{align}
& \nu_{aa} \widetilde{C}_{aa} (\widetilde{f}_a, \widetilde{f}_a) + \nu_{ab} {}^{1}\widetilde{C}_{ab} (\widetilde{f}_{a1}) + \Omega_a \frac{\partial \widetilde{f}_{a1}}{\partial \xi} = \nonumber \\ & \bu \cdot \left[ \left(\frac{\nabla p_a}{p_a} - \frac{q_a \bE}{T_a} \right) + \left( \frac{u^2}{u_{th,a}^2} - \frac{5}{2}\right) \frac{\nabla T_a}{T_a} - \frac{u_{th,a}^3}{u^3} \frac{\nu_{ab} \bu_{b}}{u_{th,a}^2} \right] \widetilde{f}_{a0}.
\end{align}
Therefore, if the ratios $\Omega_a / \nu_{ab}$ and $\nu_{aa} / \nu_{ab}$ are the same in the two-ion-species case and in the electron-ion case, the solution $\widetilde{f} (u/u_{th,a})$ is also the same, and the expression for the friction force $R_{ab}$ is the same.
As long as $m_a \ll m_b$, $\nu_{ab} / \nu_{aa} = \sqrt{2} n_b Z_b^2 / (n_a Z_a^2) $.
If species $a$ are electrons, and species $b$ are ions, then $\Omega_a / \nu_{ab} = \omega \tau$, $\nu_{ab} / \nu_{aa} = \sqrt{2} n_b Z_b^2 / n_a = \sqrt{2} Z_b$. Usually (e.g. in Braginskii\cite{Braginskii1965} or in Epperlein and Haines\cite{Epperlein1986}) transport coefficients are found in terms of $\omega \tau$ and $Z$. 

Therefore, the expression for the ion-ion friction force between light species $a$ and heavy species $b$ is the same as the expression for the electron-ion friction force in single-ion-species plasma up to substitutions $\Omega_a / \nu_{ab} \to \omega \tau $ and $\widetilde{Z} = \nu_{ab} / (\sqrt{2} \nu_{aa}) = n_b Z_b^2 / (n_a Z_a^2)$.

\end{document}